\documentclass{jpp}
\usepackage{graphicx}
\usepackage{xcolor}
\usepackage[utf8]{inputenc}
\usepackage[T1]{fontenc}
\usepackage{amsmath}
\usepackage{algorithm}
\usepackage{algpseudocode}
\usepackage{soul}
\usepackage{subcaption}
\usepackage{booktabs}   
\usepackage{siunitx}   
\usepackage{hyperref}
\usepackage{url}

\algrenewcommand\algorithmicrequire{\textbf{Input:}}
\algrenewcommand\algorithmicensure{\textbf{Output:}}

\shorttitle{Exponential Spectral Scaling}
\shortauthor{B. Jang, R. Conlin, M. Landreman}

\title{Exponential Spectral Scaling: Robust and Efficient Stellarator Boundary Optimization via Mode-Dependent Scaling}

\author{Byoungchan Jang\aff{1}
  \corresp{\email{byoungj@umd.edu}}, Rory Conlin\aff{1}, \and
  Matt Landreman\aff{1}}

\affiliation{\aff{1}Institute for Research in
Electronics and Applied Physics, University of Maryland, College Park, MD, 20742, USA}

\begin{document}

\maketitle

\begin{abstract}
Stellarator boundary optimization faces a fundamental numerical challenge: the extreme disparity between low- and high-mode amplitudes creates an optimization landscape in which direct full-spectrum approaches typically converge to poor local minima. Traditionally, this challenge has been addressed through a computationally expensive, multi-step  Fourier continuation, in which low Fourier modes are optimized first, followed by the gradual incorporation of higher modes. We present Exponential Spectral Scaling (ESS), a technique that applies a mode-dependent exponential scaling factor to each Fourier mode. Our primary implementation uses the $L_{\infty}$ norm to determine the scaling pattern, creating a square spectral decay profile that effectively reduces the dynamic range of optimization variables from 6–7 orders of magnitude to 2–3. This scaling aligns with the natural spectral decay of physically meaningful configurations and enables direct single-step optimization using the full-spectrum of boundary Fourier modes. ESS eliminates arbitrary staging decisions and reduces computation time by a factor of $\sim 2\text{–}5$ in benchmark cases. In addition to accelerating optimization, ESS improves robustness, reducing sensitivity to initial conditions, and increasing confidence in avoiding local optima. We demonstrate the effectiveness of ESS across both quasi-axisymmetric (QA) and quasi-helically symmetric (QH) configurations, using two distinct optimization toolkits: \textsc{simsopt} and \textsc{desc}.
\end{abstract}

\section{Introduction}
Stellarators present a promising pathway to steady-state magnetic confinement fusion, offering inherent stability and low recirculating power requirements due to their reliance on externally generated magnetic fields. Unlike the axisymmetric design of tokamaks, stellarators are not constrained by continuous geometric symmetry. This freedom enables the engineering of complex, fully three-dimensional magnetic fields, broadening the range of possible configurations and allowing precise tailoring of magnetic geometry to enhance performance. While this geometric flexibility is a central advantage, it also renders confinement highly sensitive to shape; without deliberate and careful design, stellarators are susceptible to increased orbit losses and degraded transport. As a result, boundary-shape optimization plays a critical role in realizing the potential benefits of stellarator-based fusion devices.

\begin{figure}
    \centering
    \includegraphics[width=1.0\textwidth]{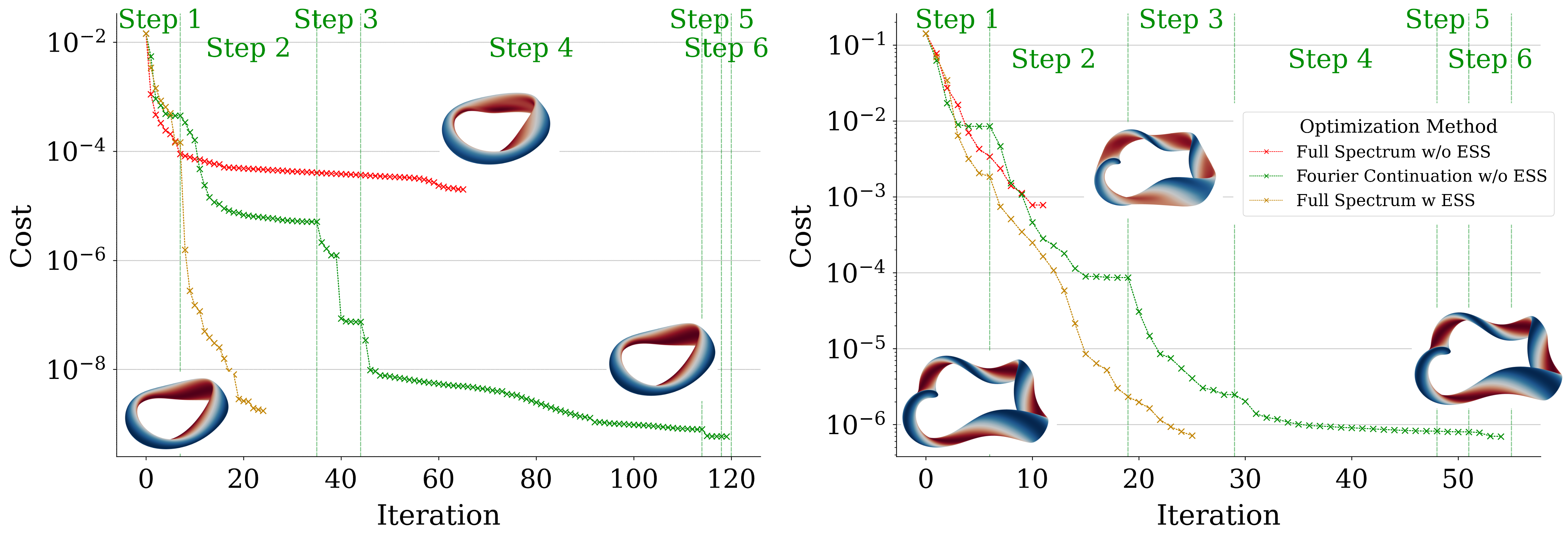}
    \caption{QA (left) and QH (right) optimization loss history comparison and final 3D stellarator configuration. The red line shows the loss history with full-spectrum using Moore's Jacobian scaling as described in Appendix~\ref{app:jacobian-scaling}, the green line shows the loss history with Fourier continuation using Moore's Jacobian scaling, and the gold line shows the loss history with full-spectrum using ESS.}
    \label{fig:figs/qs_optimization_comparison.png}
\end{figure}

Stellarator optimization traditionally separates the problem into (i) optimizing the plasma boundary shape \citep[]{nuhrenberg1988quasi, grieger1992physics, mynick2006transport, bader2019stellarator, landreman2022magnetic} and (ii) designing coils that reproduce that boundary shape~\citep{merkel1987solution, drevlak1998automated, boozer2000optimization, pomphrey2001innovations, zhu2018designing}. We focus exclusively on the first stage. For the first stage, the conventional approach often includes Fourier continuation~\citep{landreman2021stellarator}, where low-order Fourier modes are optimized first before gradually introducing higher ones. While this staged approach aids convergence, it is computationally inefficient, often over-optimizes low modes, and depends on arbitrary hyperparameters that can create artificial convergence plateaus as shown by the green loss history from Figure~\ref{fig:figs/qs_optimization_comparison.png}. In contrast, the full-spectrum approach refers to optimizing all Fourier modes simultaneously from the outset, without any staged mode introduction. Despite its conceptual simplicity, the full-spectrum approach is rarely adopted in practice because the high-dimensional parameter space can lead to slow convergence, poor optimization performance, and an increased likelihood of getting trapped in local minima as shown by the red loss history in Figure~\ref{fig:figs/qs_optimization_comparison.png}.

In this paper, we introduce Exponential Spectral Scaling (ESS) as a direct method to address the mode amplitude disparity (MAD). ESS applies a mode-specific exponential scaling that significantly reduces MAD and improves the overall optimization landscape. We evaluate ESS with different norm choices ($L_1$, $L_2$, and $L_{\infty}$), demonstrating robust convergence across various stellarator configurations. While this work focuses exclusively on local optimization, we expect ESS to be broadly compatible with global optimization frameworks and may improve their efficiency as well. In particular, because any global optimization algorithm requires box constraints, ESS also informs how to choose meaningful bounds in spectral space. As shown by the gold loss history line in Figure~\ref{fig:figs/qs_optimization_comparison.png} (which will be described in more detail in Section~\ref{sec:optimization_results}), ESS consistently delivers faster and more reliable convergence within a general trust-region framework, while avoiding nonphysical outcomes such as self-intersecting surfaces.

In various optimization settings, imbalances between low- and high-spectral
content—formalized here as MAD—have been previously noted.  In stellarator research, Fourier continuation~\citep{landreman2021stellarator} is typically used to mitigate such imbalance; penalization of high-mode-number amplitudes has also been used~\citep{singh2020optimization}.  In
aerospace inverse design, gradient-limiting Sobolev filters or staged Fourier optimization is utilized to damp high-order
shape perturbations and stabilize convergence
\citep{masters2017multilevel,kedward2020gradient}.  Geophysical full-waveform inversion confronts
analogous ``cycle-skipping'' local minima by introducing multiscale frequency
sweeps and convex misfit functions
\citep{bunks1995multiscale,sirgue2004efficient,pladys2021cycle}.  Within
machine learning, \emph{spectral bias} and \emph{F-principle} reveal that
networks learn low-frequency components first, inspiring frequency-aware loss
functions and rescaling schemes
\citep{rahaman2019spectral,xu2019frequency,wang2024multi}.  These convergent insights across
disciplines motivate our ESS as a principled
physics-agnostic remedy for MAD.

\section{Towards Reducing the Mode Amplitude Disparity}\label{sec:reducing_MAD}
\subsection{Spectral Representation of the Plasma Boundary}
Stellarator boundary shapes are typically represented using truncated double Fourier series in cylindrical coordinates. Most stellarator optimization to date has used the VMEC representation, in which \( R \) and \( Z \) are expanded as cosine and sine series in the form \( \cos(m\theta - n N_{\mathrm{FP}}\phi) \). In this paper, we instead adopt the related but distinct DESC representation defined in Equations~\eqref{eq:BoundaryR}–\eqref{eq:BoundaryZ}.
\begin{align}
    R(\theta,\zeta) &= \sum_{m=-M}^{M}\sum_{n=-N}^{N} R_n^m \, \mathcal{G}_n^m(\theta,\zeta),\label{eq:BoundaryR}\\[1mm] 
    Z(\theta,\zeta) &= \sum_{m=-M}^{M}\sum_{n=-N}^{N} Z_n^m \, \mathcal{G}_n^m(\theta,\zeta), \label{eq:BoundaryZ}
\end{align}
where $\theta\in [0,2\pi)$, the normalized toroidal angle is defined as $\zeta=\phi/N_{\mathrm{FP}}\in[0,2\pi)$, and the basis functions $\mathcal{G}_n^m(\theta,\zeta)$ are defined piecewise according to the signs of $m$ and $n$:
\begin{equation}
    \mathcal{G}_n^m(\theta,\zeta)=
    \begin{cases}
        \cos(|m|\theta)\cos(|n|\zeta), & m,n\geq 0\\
        \cos(|m|\theta)\sin(|n|\zeta), & m\geq 0,n<0\\
        \sin(|m|\theta)\cos(|n|\zeta), & m<0,n\geq 0\\
        \sin(|m|\theta)\sin(|n|\zeta), & m,n<0.
    \end{cases}
    \label{eq:doublefourier}
\end{equation}
However, this representation leads to a wide variation in the magnitude of Fourier coefficients. This effect, which we refer to as \emph{mode amplitude disparity (MAD)}, has significant consequences for optimization.

\subsection{Numerical Challenges from Mode Amplitude Disparity}
\begin{figure}
    \centering
    \subfloat{%
        \includegraphics[width=0.5\textwidth]{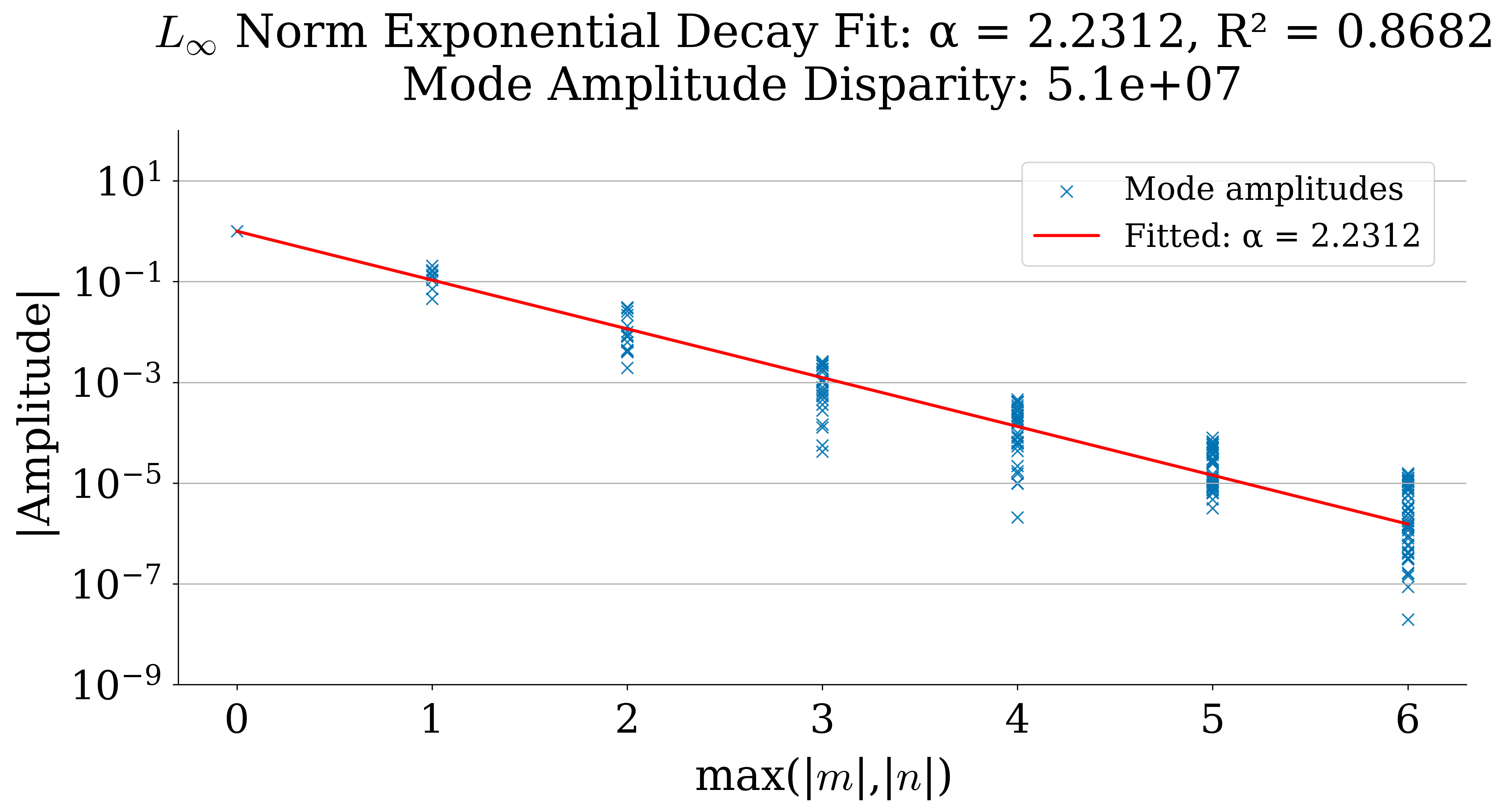}%
    }
    \hfill
    \subfloat{%
        \includegraphics[width=0.5\textwidth]{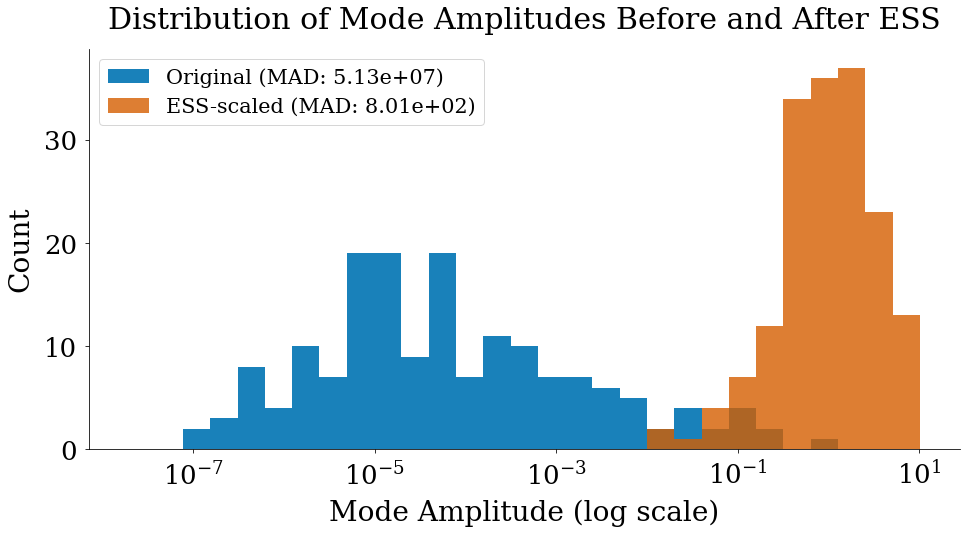}%
    }
    \caption{Mode amplitudes showing exponential decay for the LandremanPaul QA stellarator~\citep{landreman2022magnetic} (left). Distribution of mode amplitudes before and after ESS (right).}
    \label{fig:figs/qa_amplitude_decay}
\end{figure}

MAD refers to the dynamic range among the Fourier coefficients used to represent stellarator boundaries. To quantify this imbalance, we define the MAD as
\begin{align}
    \mathrm{MAD} = \frac{\max_{(m,n)} |A_n^m|}{\min_{(m,n) \,:\, A_n^m \neq 0} |A_n^m|},
\end{align}
where \(A_n^m\) is either $R^m_n$ or $Z^m_n$ from~\eqref{eq:BoundaryR}-\eqref{eq:BoundaryZ}. This metric captures the effective dynamic range across all modes, excluding identically vanishing coefficients, and highlights the challenges posed by extreme scale separation in spectral representations. This imbalance poses serious numerical challenges because optimization algorithms are sensitive to the scale of variables. When parameters differ significantly in magnitude, the resulting subproblems are poorly conditioned, leading to instability, slow convergence, or convergence to suboptimal minima~\citep[Sec.~2.2, p.~26]{nocedal2006}. As shown in Figure~\ref{fig:figs/qa_amplitude_decay} (left), the Fourier mode amplitudes for a typical configuration exhibit near-exponential decay, which is typical of spectral representations (See Appendix~\ref{app:fourier-decay} for more detail). These issues motivate the need for strategies that reduce the amplitude disparity while preserving geometric fidelity.

\subsection{Fourier Continuation as a Mitigation Strategy}
Stellarator boundary optimization seeks to minimize an objective function \( f(\mathbf{x}) \) that quantifies how far a magnetic configuration deviates from desired physical properties. Given an initial boundary shape \( \partial\Omega^{\text{init}} \), a pressure profile \( p(\psi) \), and a toroidal current profile \( I_T(\psi) \), the optimization iteratively adjusts Fourier coefficients \( \mathbf{x} = [R_n^m, Z_n^m] \) until convergence. One conventional strategy for addressing the mode amplitude disparity (MAD) in this process is \emph{Fourier continuation} (FC)—a staged approach that incrementally increases the maximum Fourier mode number from an initial low value \( M_0 \) to a target \( M_{\text{max}} \). At each stage, the current solution initializes the next, introducing additional higher-order modes progressively. This allows the optimizer to start with a truncated, stable spectral representation and gradually refine it, as detailed in Algorithm~\ref{alg:fc}.

The rationale behind this approach is twofold: (1) low-order modes tend to dominate the global structure of the plasma boundary and are easier to optimize first; and (2) introducing high-order modes incrementally avoids overwhelming the optimizer with ill-conditioned variables from the outset. This mitigates numerical instabilities and helps avoid convergence to poor local minima in early iterations. Beyond these theoretical advantages, Fourier continuation has proven effective in practice, consistently improving optimization robustness across a range of configurations.

\begin{algorithm}
\caption{Stellarator Boundary Optimization with Fourier Continuation}\label{alg:fc}
\begin{algorithmic}[1]
\Require Initial low mode $M_0$, max mode $M_{\max}$, initial boundary $\partial\Omega^{init}$, and tolerances $\{\epsilon_x, \epsilon_f, \epsilon_g\}$
\Ensure Optimized boundary $\partial\Omega$
\State $\partial\Omega \gets \partial\Omega^{init}$ \Comment{Initialize boundary}
\For{$M \gets M_0$ to $M_{\max}$} \Comment{Increase Fourier modes}
    \State Define $\mathbf{x} = [R_n^m, Z_n^m]$ for all $|m|,|n| \leq M$
    \State Compute MHD equilibrium field $B(\Omega)$ using $\partial\Omega$, $p(\psi)$, and $I_T(\psi)$
    \State Evaluate objective $f(\mathbf{x})$ from $B(\Omega)$
    \While{$\|\Delta\mathbf{x}\|_2>\epsilon_x \,\land\, |\Delta f|>\epsilon_f \,\land\, \|\nabla f(\mathbf{x})\|_2>\epsilon_g$}\Comment{Convergence check on objective}
        \State Update $\partial\Omega$ using an optimization routine
        \State Compute updated $B(\Omega)$ from new $\partial\Omega$, $p(\psi)$, and $I_T(\psi)$
        \State Re-evaluate $f(\mathbf{x})$ from the updated $B(\Omega)$
    \EndWhile
    \State $\partial\Omega^{init} \gets \partial\Omega$ \Comment{Propagate optimized boundary}
\EndFor
\State \Return $\partial\Omega$
\end{algorithmic}
\end{algorithm}

Although widely used, Fourier continuation relies on heuristic choices that are not necessarily optimal. The typical approach adds modes incrementally by bounding $\|m\|$ and $\|n\|$ at each step, but this specific ordering is arbitrary and may not align with the most relevant directions in parameter space. Such rigid heuristics can slow convergence or bias solutions toward subspaces favored by early mode choices.

\section{Exponential Spectral Scaling (ESS) Methodology}
Having outlined the challenges posed by mode amplitude disparities and the limitations of existing mitigation strategies, we now introduce Exponential Spectral Scaling (ESS) - a principled method for rebalancing spectral coefficients prior to optimization. This section presents the mathematical formulation of ESS, describes how it integrates into the optimization workflow, and evaluates its effects on numerical performance and shape evolution.
We begin by denoting the original Fourier coefficients representing the stellarator boundary as
\begin{align}
    \mathbf{x}_{\text{original}} = [R_n^m,\, Z_n^m].
\end{align}
The ESS transformation is defined by
\begin{align}
\mathbf{x}_{\text{scaled}} = 
  \frac{\mathbf{x}_{\text{original}}}{\exp\!\left(-\alpha\, g(m,n)\right)}.
\end{align}
where $\alpha\ge0$ is a user-specified decay parameter that controls the degree of scaling, and $g(m,n)$ is mode norm function that assigns relative importance to each spectral mode indexed by $(m,n)$. We consider the following choices for $g(m,n)$: $g(m,n) = |m|+|n|$ produces a diamond-shaped profile (the $L_1$ norm), $g(m,n) = \sqrt{m^2+n^2}$ gives a circular profile (the $L_2$ norm), $g(m,n) = \max\{|m|,|n|\}$ results in a square profile (the $L_{\infty}$ norm). Each of these choices defines a distinct decay geometry in Fourier space. The exponential scaling preferentially reduces the magnitude of high-order modes, thereby compressing the dynamic range of $x_{\text{original}}$ from 6–7 orders of magnitude to approximately 2–3 as shown in Figure~\ref{fig:figs/qa_amplitude_decay}, thereby improving the conditioning of the optimization problem.

\begin{figure}
    \centering
    \includegraphics[width=0.55\textwidth]{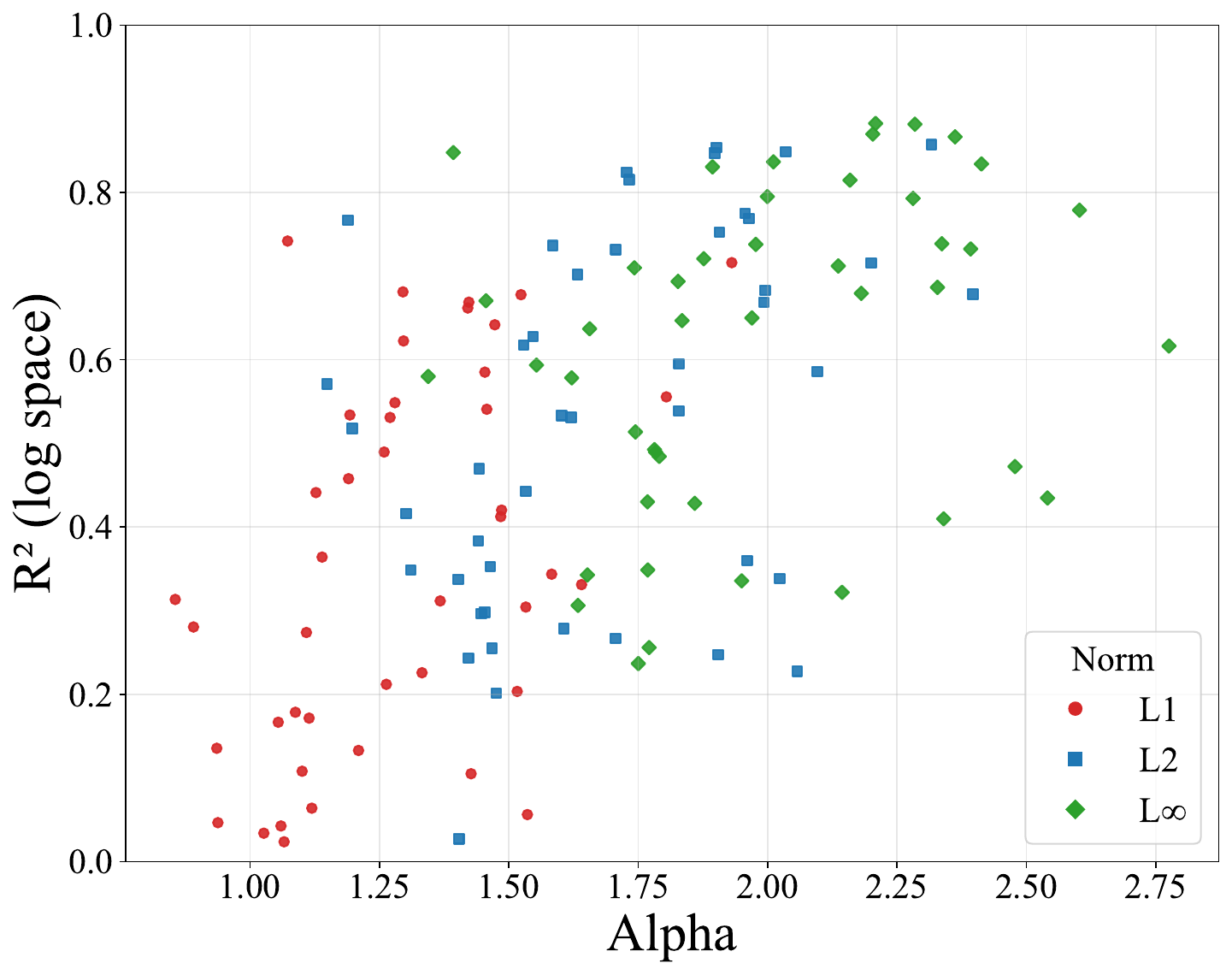}
    \caption{Fitted decay rate \(\alpha\) versus coefficient of determination \(R^2\) across a survey of  stellarators. For \(L_2\) and \(L_\infty\) decay geometries, \(\alpha\) typically falls in the range \(1.1\!-\!2.8\); for \(L_1\), the range is narrower (\(0.7 \le \alpha \le 1.9\)) and fit quality is generally lower (\(-0.3 \le R^2 \le 0.77\)).}
    \label{fig:rsquared_vs_alpha}
\end{figure}

To calibrate a practical decay rate, we analyzed a library of  stellarator configurations drawn from~\cite{kappel2024magnetic}. Exponential decays of the form $\exp(-\alpha g(m,n))$ were fitted to the Fourier coefficients, using $\log(|A^m_n|)$ in the regression and reporting $R^2$ values to quantify fit quality; Figure~\ref{fig:figs/qa_amplitude_decay} shows one representative example.  As shown in Figure~\ref{fig:rsquared_vs_alpha}, for $L_2$ and $L_\infty$, the fitted decay rates $\alpha$ cluster between about 1.1 and 2.8, while for $L_1$ the range is narrower and the fits are systematically worse. A likely reason is that stage-1 optimizations with Fourier continuation are typically based on square domains in $(m,n)$ space, which are more naturally aligned with $L_\infty$ or $L_2$ geometries than with $L_1$. The main takeaway from this survey is that reasonable choices of $\alpha$ fall in the range $\sim1$–3, providing both intuition for setting default values and justification for using $L_\infty$ as the norm in subsequent experiments. 

Building on the formulation above, ESS is applied as a preconditioner to the Fourier coefficient vector $x_{\text{original}}$ prior to optimization. Once scaled, the optimization proceeds over $x_{\text{scaled}}$ using a standard algorithm (e.g., trust-region method). In practice, ESS can be conveniently implemented by setting the \texttt{x\_scale} keyword argument in \texttt{scipy.optimize.least\_squares}. Algorithm~\ref{alg:ess} outlines the ESS-based optimization procedure without the loop over $M$ from Algorithm~\ref{alg:fc}.

\begin{algorithm}
\caption{Stellarator Boundary Optimization with ESS}\label{alg:ess}
\begin{algorithmic}[1]
\Require Initial boundary $\partial\Omega^{\text{init}}$, full set of Fourier coefficients $\mathbf{x}=[R_n^m, Z_n^m]$, convergence tolerance $\{\epsilon_x, \epsilon_f, \epsilon_g\}$, decay parameter $\alpha$, and norm function $g(m,n)$
\Ensure Optimized boundary $\partial\Omega$
\State Compute the pre-conditioned coefficients:
\[
\mathbf{x}_{\text{scaled}}
  = \frac{\mathbf{x}_{\text{original}}}
         {\exp\!\bigl(-\alpha\, g(m,n)\bigr)}
\]
\State Compute the MHD equilibrium field $B(\Omega)$ using $\partial\Omega^{\text{init}}$, the pressure profile $p(\psi)$, and the current profile $I_T(\psi)$
\State Evaluate the objective function
\[
f\!\bigl(\mathbf{x}_{\text{scaled}}
        \exp\!\bigl(-\alpha\, g(m,n)\bigr)\bigr)
\]
\While{$\|\Delta\mathbf{x}\|_2>\epsilon_x\,\land\,
       |\Delta f|>\epsilon_f\,\land\,
       \|\nabla f(\mathbf{x})\|_2>\epsilon_g$}
    \State Update $\partial\Omega$ using the trust-region algorithm applied to $\mathbf{x}_{\text{scaled}}$
    \State Recompute $B(\Omega)$ and update
          $f\!\bigl(\mathbf{x}_{\text{scaled}}
          \exp\!\bigl(-\alpha\, g(m,n)\bigr)\bigr)$
\EndWhile
\State \Return $\partial\Omega$
\end{algorithmic}
\end{algorithm}

As a result, the optimizer is able to take more meaningful steps early in the optimization process, without becoming trapped in ill-conditioned subspaces. We hypothesize that the balanced scaling results in a smoother optimization landscape, effectively serving as a continuous analogue to Fourier continuation. By enabling direct full-spectrum optimization, ESS obviates the need for arbitrary staging decisions and the staged mode activation required by traditional Fourier continuation methods. This reduces computational overhead and allows the optimizer to operate on a well-conditioned control space.

High-order Fourier modes are often associated with fine-scale features on the plasma boundary that are difficult to resolve numerically and may not correspond to physically meaningful structures. If left unconstrained, these modes can introduce artificial oscillations, kinks, or even self-intersections in the boundary geometry. ESS reduces the influence of these high-order components early in the optimization process by attenuating their amplitudes. This prioritizes smooth, well-resolved shape components. Importantly, ESS does not remove high-order modes; it simply ensures that their effect is proportionate to their representational significance. In Section~\ref{sec:experiments}, we present empirical evidence showing that ESS results in a smoother and faster convergence without instances of pathological geometries.

\section{Numerical Experiment Results}\label{sec:experiments}

We evaluate the performance of ESS on two benchmark stellarator configurations: LandremanPaul QA and QH \citep{landreman2022magnetic}. Our experimental comparisons focus on convergence dynamics, computational efficiency, and the quality of the resulting plasma boundary shapes.

\subsection{Optimization Problem}\label{sec:optimization_problems}
The optimization problem solved in \citet{landreman2022magnetic} determines a stellarator boundary that minimizes quasisymmetry errors while meeting the target rotational transform and aspect ratio goals. The independent variables are the Fourier coefficients of the plasma boundary represented in section 2, truncated to a maximum poloidal and toroidal mode number of \( m_{\max} = n_{\max} = 6 \).  These independent variables collect into the vector $\mathbf{x} = [R_n^m, Z_n^m]$. The total cost is a weighted sum of three terms, 
\begin{align}
    f(\mathbf x)= w_{\mathrm{QS}}f_{\mathrm{QS}}+ w_{\bar\iota}f_{\bar\iota}+ w_{A}f_{A}, \label{eq:total_cost} 
\end{align}
where $w_{\mathrm{QS}}$, $w_{\bar\iota}$ and $w_{A}$ are user--chosen positive weights, all chosen to be 1 in this study. Following \cite{landreman2022magnetic}, the error in quasisymmetry is measured by
\begin{align}
    f_{\mathrm{QS}} = \int_V \biggl[ \frac{1}{B^{3}} \Bigl( (N - \iota M)\, \mathbf{B} \times \nabla B \cdot \nabla \psi - (M G + N I)\, \mathbf{B} \cdot \nabla B \Bigr) \biggr]^2 \, \mathrm{d}^3x, \label{eq:f_qs}
\end{align}
where the integral is taken over the plasma volume $V$. Here, $M$ and $N$ are the integers defining the desired symmetry type, $G(\psi)$ is $\mu_0/(2\pi)$ times the poloidal current outside the surface, and $I(\psi)$ is $\mu_0/(2\pi)$ times the toroidal current inside the surface. For all computations in this work, $M = 1$ and $I = 0$, though we retain the general form of the equation for completeness.
Instead of penalizing the difference between the realized and target rotational transforms \emph{on every surface}, we use a single mean $\bar{\iota} = \int_{0}^{1} ds \,\iota\, $, and define 
\begin{equation} 
    f_{\bar\iota}=\bigl[\bar\iota-\bar\iota_{\mathrm{tar}}\bigr]^{2},\quad \bar{\iota}_{\mathrm{tar}} \; \text{is a specified target value}.\label{eq:f_iota} 
\end{equation}
Lastly, the aspect ratio is $A=R/a$, so that 
\begin{equation} 
    f_{A}=\bigl[A(\mathbf x)-A_{\mathrm{tar}}\bigr]^{2}. \label{eq:f_A} 
\end{equation} 
Throughout this work we adopt the target values $(\iota_{\text{tar}},A_{\text{tar}})=(0.42,\,6.0)$ for the quasi-axisymmetric (QA, $N_{\mathrm{FP}}=2$) configuration and $(\iota_{\text{tar}},A_{\text{tar}})=(1.24,\,8.0)$ for the quasi-helically symmetric (QH, $N_{\mathrm{FP}}=4$) configuration. Except for Figure~\ref{fig:figs/figure16_objective_history.pdf} showing the efficacy of ESS in \textsc{simsopt}~\citep{landreman2021simsopt}, optimizations were carried out with the \textsc{desc} code~\citep{panici2023desc}. The optimization was performed using the trust-region reflective (TRF) algorithm, as implemented in \texttt{scipy.optimize.least\_squares} with \texttt{method='trf'}. Comparable results were also obtained using the Levenberg–Marquardt (LM) algorithm from \textsc{MINPACK}. Initial conditions are based on rotating elliptical boundary shapes.

\subsection{Stellarator Optimization Results}\label{sec:optimization_results}

\begin{figure}
    \centering
    \begin{minipage}{.32\linewidth}
            \begin{subfigure}[t]{1.0\linewidth}
                \includegraphics[width=\textwidth]{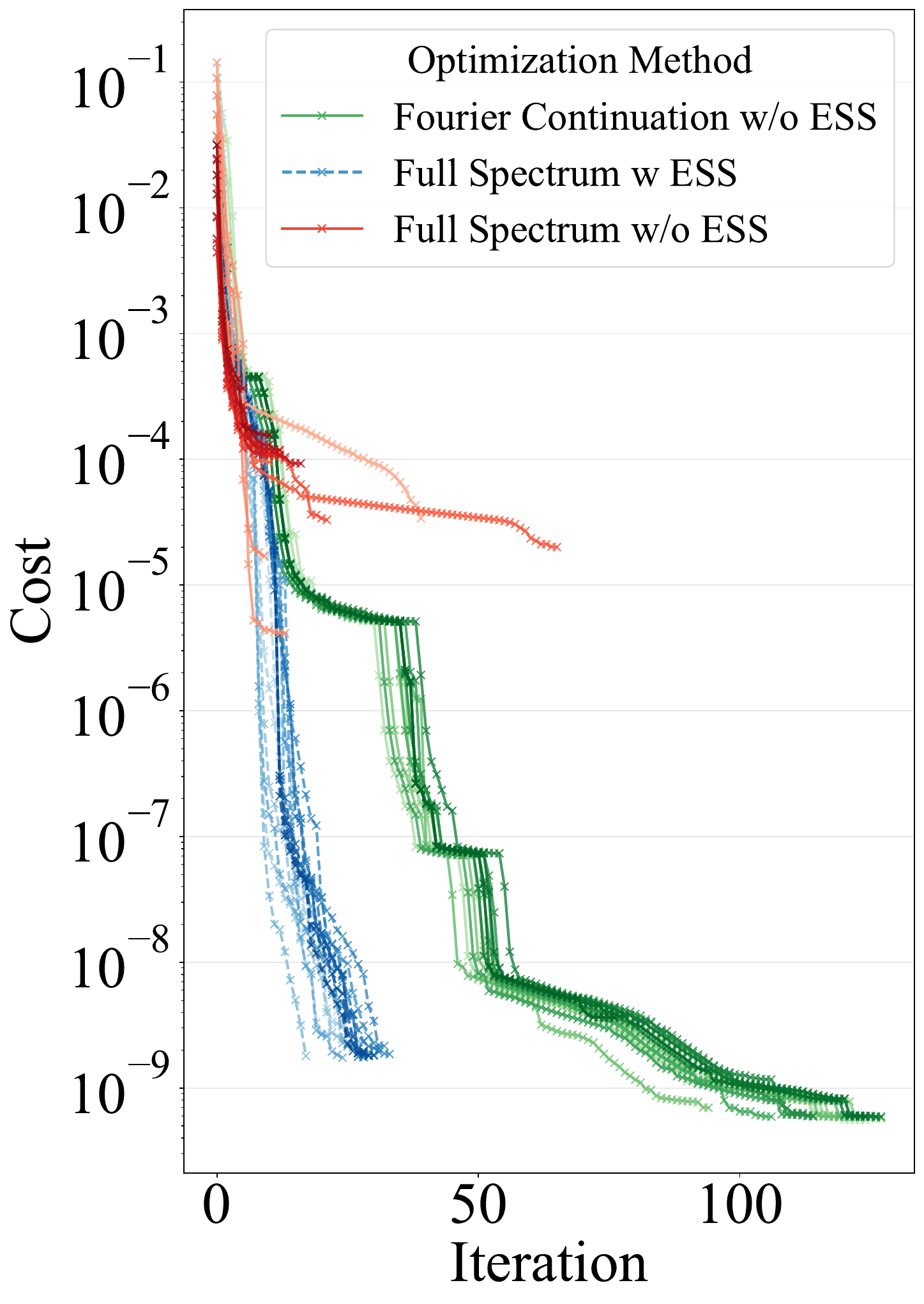}
                \caption{Loss history}
                \label{fig:ESS_vs_FC}
            \end{subfigure}
    \end{minipage}
    \hspace{0.01\linewidth}  
    \begin{minipage}{.30\linewidth}
        \begin{subfigure}[t]{1.0\linewidth}
            \includegraphics[width=\textwidth]{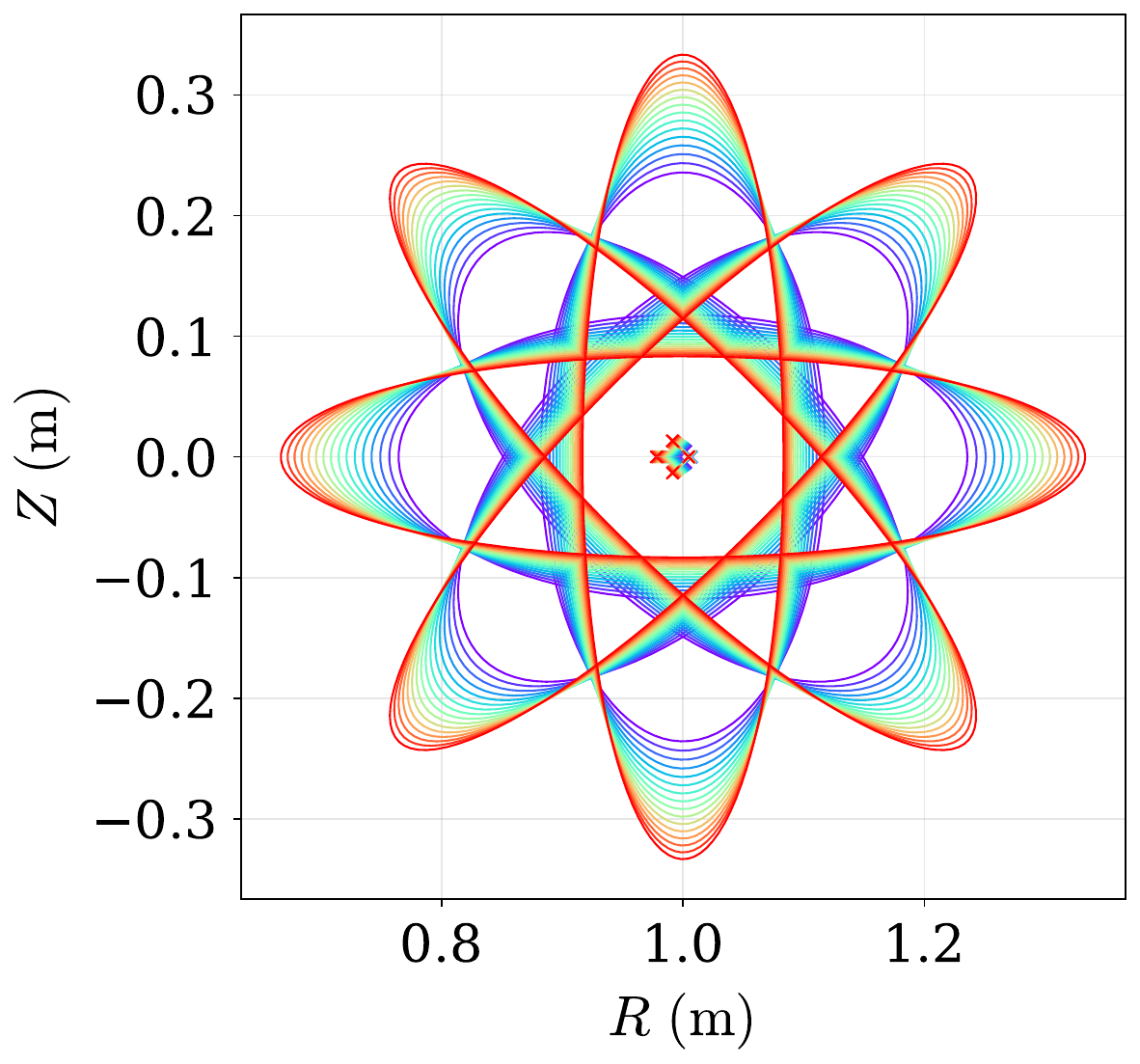}
            \caption{Initial Conditions}
            \label{fig:IC_cross-sections}
        \end{subfigure} 
        
        \vspace{0.4cm}  

        \begin{subfigure}[b]{1.0\linewidth}
            \includegraphics[width=\textwidth]{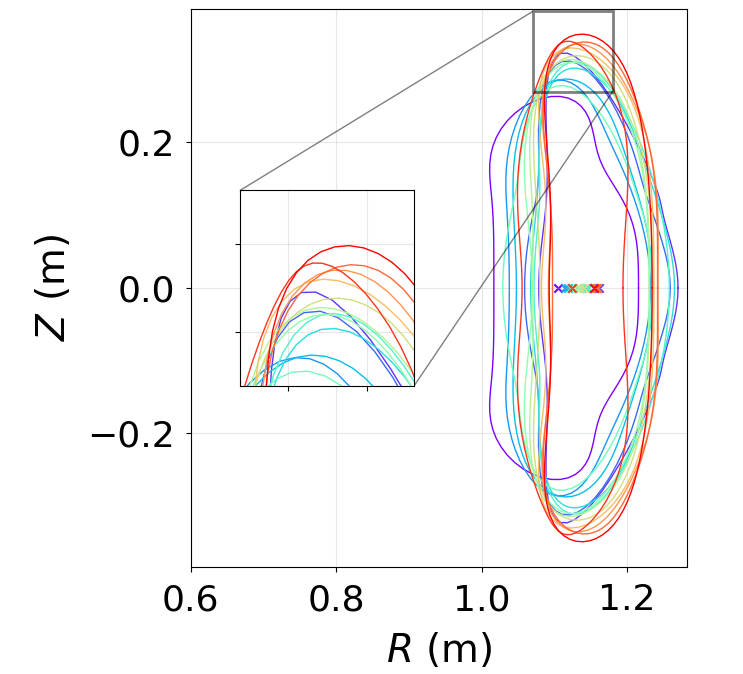}
            \caption{Full-Spectrum with Jacobian Scaling}
            \label{fig:failed_cross-sections}
        \end{subfigure} 
    \end{minipage}
    \begin{minipage}{.30\linewidth}
        \begin{subfigure}[t]{1.0\linewidth}
            \includegraphics[width=\textwidth]{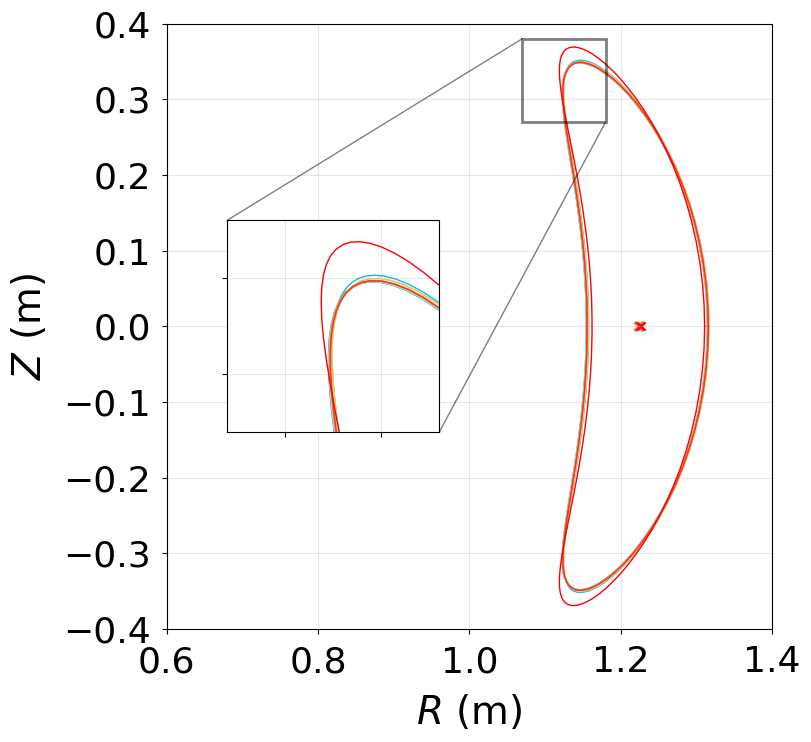}
            \caption{Fourier Continuation}
            \label{fig:FC_cross-sections}
        \end{subfigure} 
        
        \vspace{0.4cm}  

        \begin{subfigure}[b]{1.0\linewidth}
            \includegraphics[width=\textwidth]{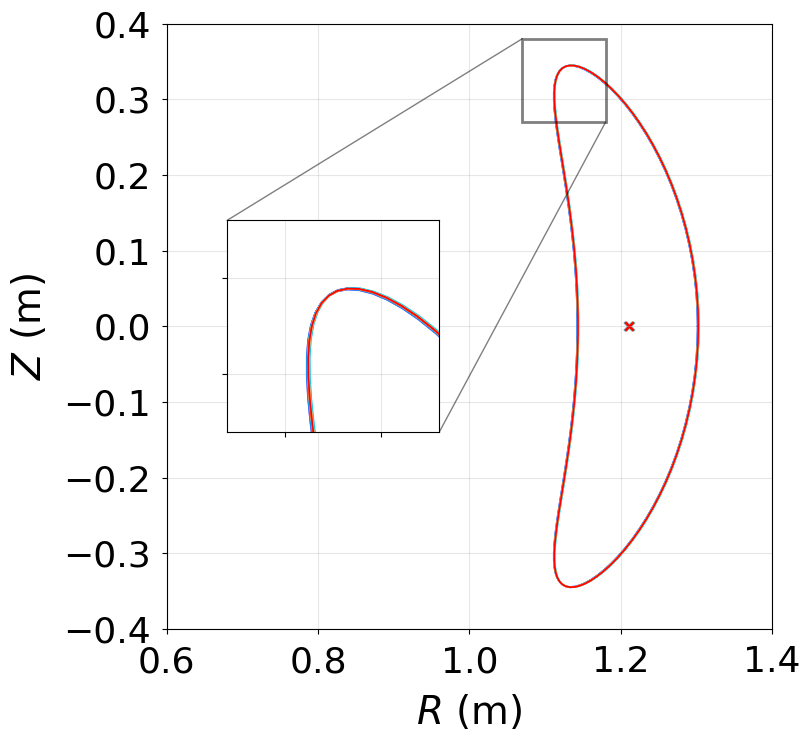}
            \caption{ESS}
            \label{fig:ESS_cross-sections}
        \end{subfigure} 
    \end{minipage}
    \caption{(a) Loss histories for three optimization methods under varying initial elongations. (b) Cross-sections of initial shape varying elongation values. (c) The stalled optimizations highlight the limitations of Jacobian scaling alone, leading to distorted shapes. (d) Final cross-sections from Fourier continuation show much less variation than in panel (c), with only minor residual differences. (e) ESS yields consistent final boundaries with minimal run-to-run variation.}
    \label{fig:convergence_qa}
\end{figure}

Figure~\ref{fig:figs/qs_optimization_comparison.png} presents the loss history during optimization, where ESS demonstrates a smoother and more rapid decrease in the objective function compared to the conventional Fourier continuation method. The vertical green lines mark the iterations at which the number of Fourier modes is increased—that is, where the \(M\) in Algorithm~\ref{alg:fc} is incremented—thus expanding the degrees of freedom in the optimization problem. For both the full-spectrum and Fourier continuation methods, Jacobian-based variable scaling is used via the \texttt{x\_scale='jac'} setting in \texttt{scipy.optimize.least\_squares}, as described in Appendix~\ref{app:jacobian-scaling}. When \texttt{x\_scale} is instead set to 1, the performance of full-spectrum optimization degrades significantly, while Fourier continuation performs similarly with or without Jacobian scaling. Figure~\ref{fig:figs/qs_optimization_comparison.png} compares the final 3D configurations for the QA and QH benchmarks obtained using ESS against those from Fourier continuation. For this figure, we employed the $L_{\infty}$ norm on $(m,n)$ with $\alpha=1.6$. The direct full-spectrum optimization enabled by ESS removes the need for arbitrary staging decisions and avoids the artificial convergence plateaus typically observed with Fourier continuation. The improved convergence and quality of the optimized configurations underscore the efficacy of ESS in stellarator boundary optimization.

Figure~\ref{fig:convergence_qa} further compares the convergence histories of ESS, FC, as well as optimizing the full-spectrum of modes with $|m|, |n| \le 6$ without FC for the QA problem across an ensemble of initial conditions varying initial elongations. Figure~\ref{fig:ESS_vs_FC} (left) shows the objective value versus iteration count for all three methods, with ESS consistently achieving faster convergence. The right panels display the final cross-sections at $\phi = 0$ with Fourier continuation and ESS, demonstrating that ESS reaches more consistent optima with less variation than FC. Figure~\ref{fig:failed_cross-sections} further illustrates that both FC and ESS resolve the issues seen in the full-spectrum case, with ESS requiring fewer iterations while FC attains slightly lower objective values. Here we employ the $L_\infty$ norm on $(m,n)$ with $\alpha = 1.6$.

\subsection{Robustness to Norm Choice and Decay Rate}

\begin{figure}
    \centering
    \subfloat{%
        \includegraphics[width=0.45\textwidth]{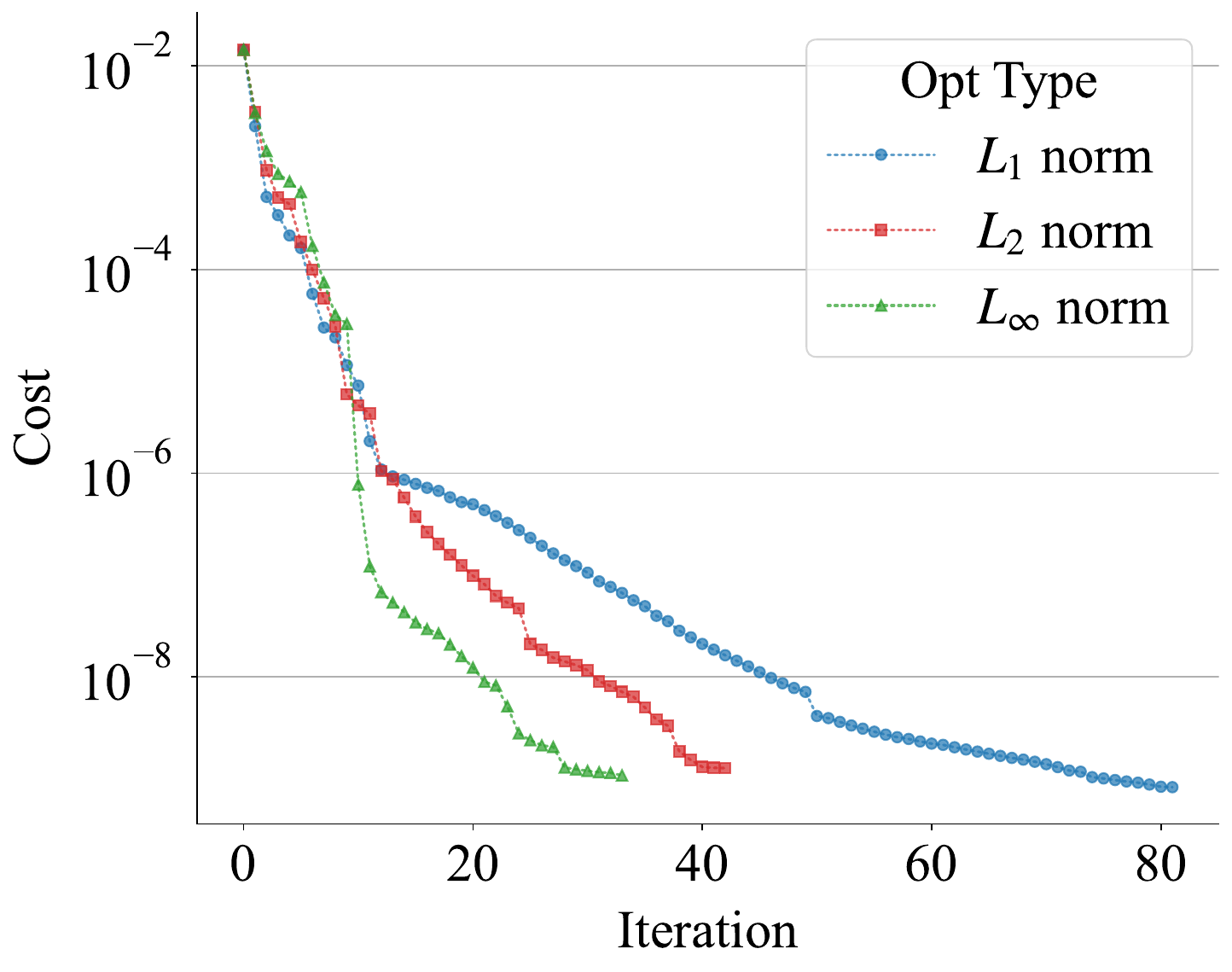}%
    }
    \hfill
    \subfloat{%
        \includegraphics[width=0.45\textwidth]{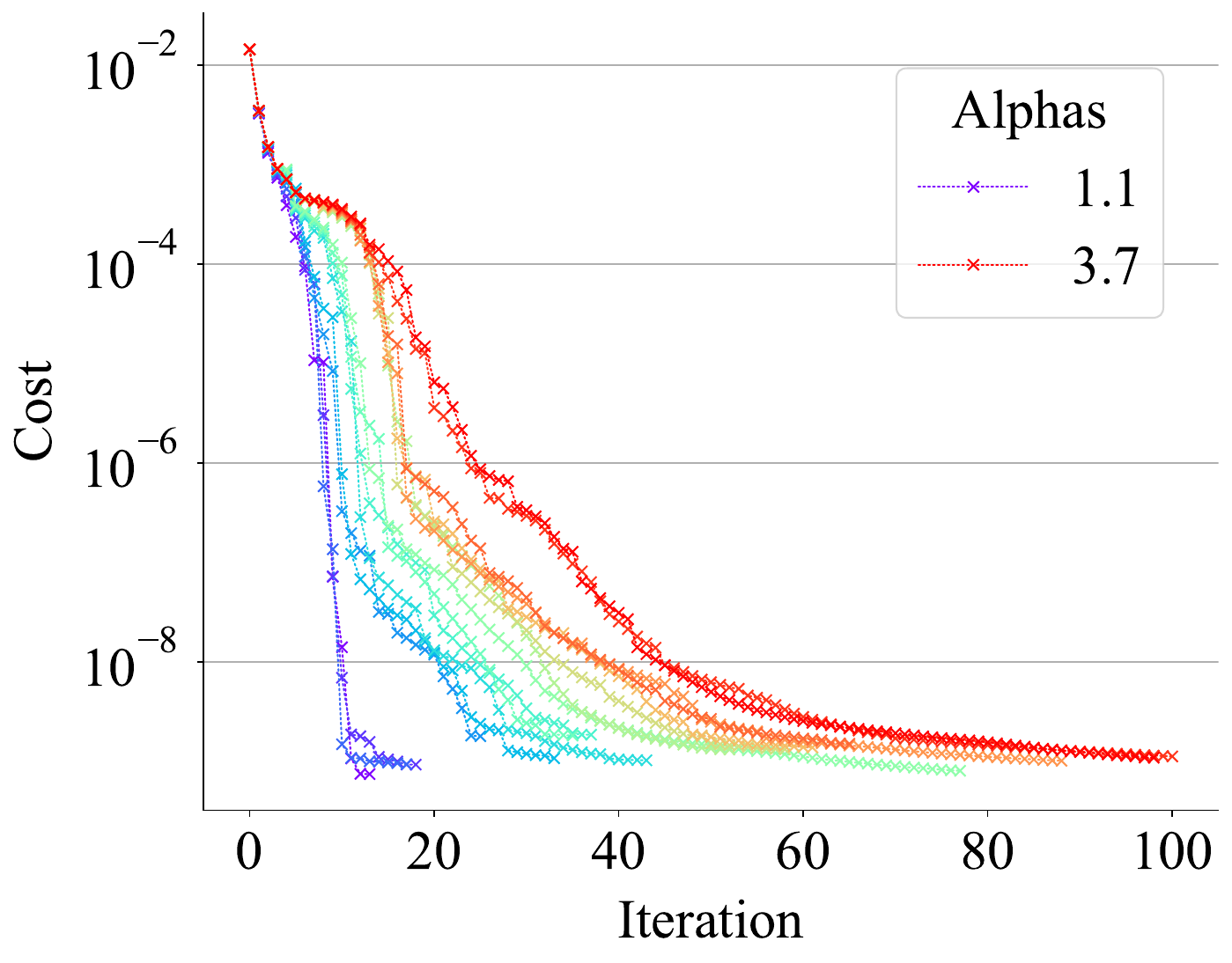}%
    }
    \caption{Loss histories for different ESS norm types for $\alpha = 1.8$ (left) and for varying $\alpha$ values under $L_\infty$ scaling (right), for the QA problem.}
    \label{fig:ess_comparison}
\end{figure}

We conducted a systematic analysis of the ESS method across different norm functions and decay parameters $\alpha$ for the QA problem. Figure~\ref{fig:ess_comparison} illustrates the convergence behavior for three norm types—$L_1$, $L_2$, and $L_{\infty}$—as well as a sweep over various $\alpha$ values. While all three norm types and the tested $\alpha$ values eventually reach similarly low objective values, some parameter choices achieve convergence more quickly.

\begin{figure}
    \centering
    \includegraphics[width=0.9\textwidth]{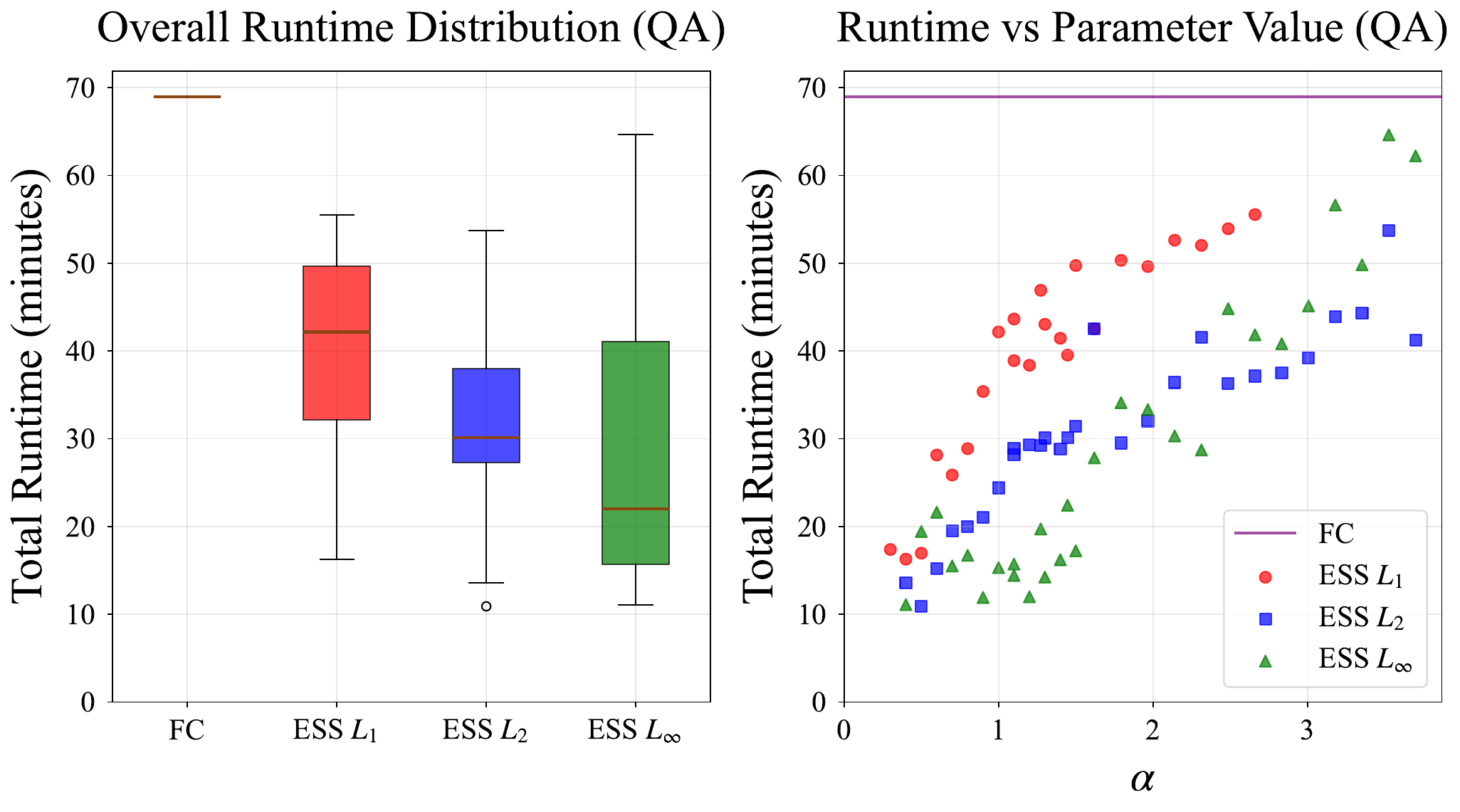}
    \caption{Wall clock time to reach an objective of $10^{-8}$ for different combinations of $\alpha$ and norm types. All the optimizations have initial elongation of $2.8$. The result from an optimization using FC is shown at a single parameter value, as FC was used only as a comparison case with the same initial elongation as the ESS runs. For ESS, $\alpha$ was varied and missing markers for ESS methods indicate optimizations in which the objective did not reach $10^{-8}$.}
    \label{fig:ess_optimal}
\end{figure}

Figure~\ref{fig:ess_optimal} and Table~\ref{table:performance_qa} further quantify robustness of ESS for the QA problem, showing the number of iterations required to reach a target objective of $10^{-8}$ for various $(\alpha, \text{norm})$ combinations. Robust convergence is observed for $\alpha$ values ranging from 0.5 to 3.3, with optimal performance typically occurring between 0.8 and 1.5. This performance-based window partially overlaps with the fitted decay rates reported in Figure~\ref{fig:rsquared_vs_alpha}, which extend to higher values, especially for $L_2$ and $L_\infty$. The difference may arise because the fitted $\alpha$ describes geometric decay in final equilibria, while the algorithmic $\alpha$ governs progress along the optimization path. The absence of a marker indicates that the objective was not reduced below $10^{-8}$, which is rare in the tested range. All ESS variants demonstrate successful convergence across a wide range of hyperparameters. In particular, the $L_2$ and $L_{\infty}$ norms yield smoother convergence and broader basins of attraction compared to the $L_1$. These norms also achieve up to a 5× reduction in wall-clock time relative to the Fourier continuation method. For consistent results, we recommend a default value of $\alpha = 1.0$ and using the $L_{\infty}$ norm.

\begin{table}
\captionsetup{justification=centering}
\caption{
Performance comparison (QA) corresponding to Figure~\ref{fig:ess_optimal}, with uncertainties reported as ±1$\sigma$
}\label{table:performance_qa}
\vspace{0.5em}
\ttfamily
\begin{tabular}{lcccccc}
\toprule
Method & Count & Total Time (min) & Opt Time (min) & Opt Steps & Final Cost \\
\midrule
FC      & 16 & $69.0 \pm 0.0$  & $68.5 \pm 0.0$  & $6$ & $5.90\text{e-}10 \pm 0.0$ \\
ESS L1  & 23 & $39.5 \pm 12.1$ & $39.1 \pm 12.1$ & $1$ & $2.04\text{e-}09 \pm 3.2\text{e-}09$ \\
ESS L2  & 28 & $31.3 \pm 10.2$ & $30.9 \pm 10.2$ & $1$ & $1.16\text{e-}09 \pm 4.8\text{e-}10$ \\
ESS Linf& 28 & $28.7 \pm 16.1$ & $28.3 \pm 16.1$ & $1$ & $1.22\text{e-}09 \pm 6.3\text{e-}10$ \\
\bottomrule
\end{tabular}
\end{table}

\subsection{Results using SIMSOPT}

\textsc{desc} is a standalone equilibrium and optimization code that solves the MHD force-balance using a spectral representation, with derivatives provided via automatic differentiation in a single coherent framework \citep{dudt2020desc}. In contrast, \textsc{simsopt} is a modular optimization environment that orchestrates external solvers—here, \textsc{vmec} for equilibrium and its post-processing tools—before assembling objective terms for optimization \citep{landreman2021simsopt}. Both codes expose the boundary Fourier coefficients $\mathbf{x}=[R_n^m, Z_n^m]$ as optimization variables, but differ in how equilibria, derivatives, and parameterizations are handled: \textsc{desc} computes analytic derivatives, whereas \textsc{simsopt} relies on finite-differencing of \textsc{vmec}. In particular, \textsc{simsopt} adopts the \textsc{vmec} convention,
\[
R(\theta,\zeta) = \sum_{m,n} R^{c}_{m,n} \cos(m\theta - n\zeta), \quad
Z(\theta,\zeta) = \sum_{m,n} Z^{s}_{m,n} \sin(m\theta - n\zeta),
\]
which follows directly from imposing stellarator symmetry [cf. eq.~\ref{eq:doublefourier}] and contrasts with the internal conventions of \textsc{desc}. The two parameterizations are related by trigonometric angle-sum formulas, so the choice only changes the coefficient representation rather than the underlying geometry. These differences affect per-iteration cost and the numerical form of the Jacobian, but not the structure of the optimization problem itself.

\begin{figure}
    \centering
    \includegraphics[width=0.8\textwidth]{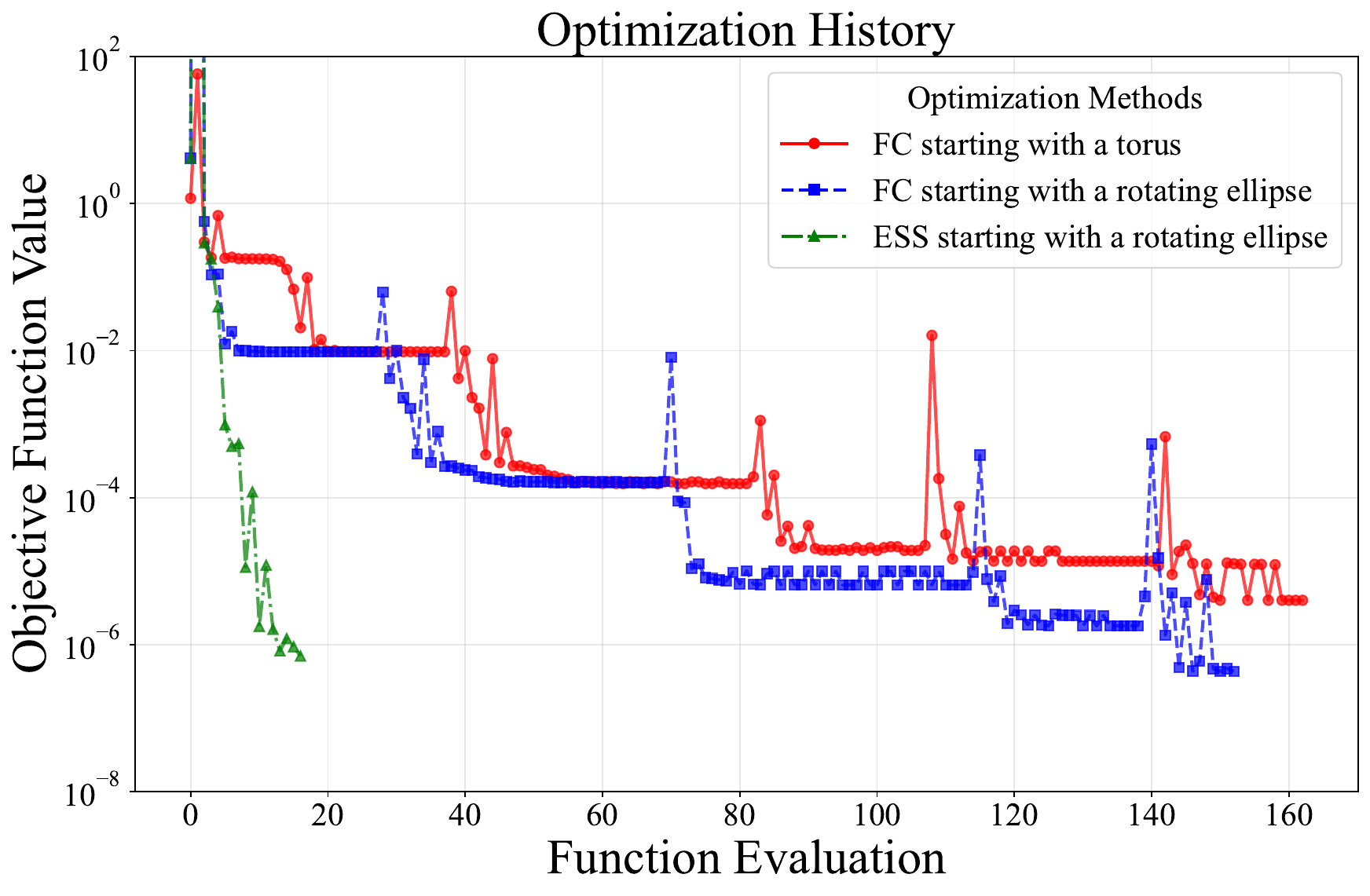}
    \caption{Loss histories for three optimization runs using \textsc{simsopt} for the same QA optimization problem discussed in Section~\ref{sec:optimization_problems}. $L_{\infty}$ is used for the norm function $g(m,n)$ and $\alpha = 1.2$ is used.} \label{fig:figs/figure16_objective_history.pdf}
\end{figure}

Despite these backend differences, ESS exhibits similar behavior in both environments. Using the same optimization problem, weights, and targets as in Section~\ref{sec:optimization_problems}, the same rotating-ellipse initial conditions, and the same trust-region reflective (TRF) algorithm via \texttt{scipy.optimize.least\_squares}, \textsc{simsopt} produces loss trajectories and final boundaries that closely mirror the \textsc{desc} results. Figure~\ref{fig:figs/figure16_objective_history.pdf} shows three \textsc{simsopt} runs for the QA case with $g(m,n)=L_{\infty}$ and $\alpha=1.2$, demonstrating the same smoothing and acceleration of convergence observed in \textsc{desc}. Any differences in iteration counts or wall time arise from equilibrium/Jacobian implementations rather than from ESS itself. Since ESS acts purely as a preconditioning in the coefficient space, it is solver-agnostic and effective with both \textsc{desc} and \textsc{simsopt}.

\section{Conclusion}
We have introduced \emph{ESS}, a single-step pre-conditioning technique that rescales the Fourier representation of stellarator boundaries to overcome the extreme mode-amplitude disparity that plagues stellarator boundary optimization workflows. By recasting the boundary-shape problem into a well-conditioned space, ESS collapses a dynamic range of six to seven orders of magnitude in Fourier coefficients to just two or three. This eliminates the need for Fourier continuation, while also achieving convergence in fewer iterations. In systematic QA trials spanning decay rates $0.0\le\alpha\le3.7$ and $L_{1}$, $L_{2}$, and $L_{\infty}$ norm geometries, this conditioning enabled direct full-spectrum optimizations that converged up to five times faster than Fourier continuation and reliably avoided non-physical local minima.

Beyond accelerating current stellarator boundary design, ESS opens promising directions for future work. First, applying the transformation to quasi-isodynamic (QI) optimizations will test whether the conditioning advantages persist outside QA/QH symmetry classes. Typically, in QI optimizations more resolution is needed in the toroidal coordinate, so ESS may need to be anisotropic in (m,n) space. Global search strategies (e.g., Bayesian optimization or other evolutionary methods) performed directly in the ESS-conditioned space may better navigate disconnected basins while avoiding the typical difficulties associated with poorly scaled parameters. Another open direction is scaling the degrees of freedom associated with the equilibrium interior during an equilibrium solve, where large disparities in sensitivity may similarly degrade conditioning. Likewise, testing algebraic decay functions in place of exponential scaling could reveal whether alternative spectral transformations provide comparable robustness with different trade-offs. 

Two targeted questions remain for stage-2 coil optimization: (i) does FC/ESS benefit single-filament inverse Biot--Savart optimizations (e.g., FOCUS-style formulations~\citep[]{zhu2018designing})? and (ii) does it aid optimization of winding pack orientation for finite-build coils? A separate and broader question is why Jacobian-based variable scaling—so effective in many nonlinear least-squares problems—performs comparatively poorly in the stage-1 context. Clarifying these issues will help determine how broadly ESS can streamline the stellarator design pipeline.

\section*{Acknowledgements}
We would like to acknowledge valuable discussions with Rahul Gaur, Nathaniel Barbour, Alan Kaptanoglu, Stefan Buller, Yigit Elmacioglu, Daniel Dudt, Misha Padidar, Andrew Giuliani, Georg Stadler, and David Bindel. This work was supported by the U.S. Department of Energy under contract No. DE-FG02-93ER54197. This research used resources of the National Energy Research Scientific Computing Center (NERSC), a U.S. Department of Energy Office of Science User Facility located at Lawrence Berkeley National Laboratory, operated under Contract No. DE-AC02-05CH11231 using NERSC award FES-ERCAP-mp217-2025.

\section*{Data availability statement}
Data associated with this study can be downloaded from 
\href{https://doi.org/10.5281/zenodo.17145011}{Zenodo (DOI: 10.5281/zenodo.17145011)} \cite{jang2025}.

\appendix

\section{Natural exponential decay of Fourier mode coefficients}\label{app:fourier-decay}

In many physical systems, especially those governed by analytic functions, the Fourier coefficients of the system decay exponentially with mode number. This explains why only a modest number of Fourier modes are needed to capture the essential geometry of typical stellarator boundaries, and motivates our design of ESS.

Mathematically, if a function \( f(\theta) \) is analytic in a strip around the real axis, its Fourier series coefficients \( \hat{f}_n \) decay exponentially:
\begin{equation}
|\hat{f}_n| \leq C e^{-\alpha |n|},
\end{equation}
for some constants \( C, \alpha > 0 \). This principle is rigorously proven in classical harmonic analysis; see, for example, \citep[Thm.~XII, p.~16]{paley1934fourier}.

This exponential decay is not merely a convenient numerical artifact but a reflection of physical regularity. Stellarator boundaries derived from equilibrium calculations or experimental measurements tend to exhibit such regularity. Thus, the use of a mode-dependent exponential scaling—rather than uniform or Jacobian-based scaling—is naturally aligned with the spectral structure of the problem and helps mitigate the large dynamic range across mode amplitudes in boundary optimization.

\section{Jacobian-Based Variable Scaling in Optimization}\label{app:jacobian-scaling}

A common approach to improving the robustness and convergence of nonlinear least squares optimization is to scale variables using the inverse norms of the Jacobian matrix's columns. This technique is particularly associated with trust-region methods and is implemented in solvers such as \texttt{MINPACK} and \texttt{scipy.optimize.least\_squares} under the option \texttt{scale='jac'}. The method originates from the work of~\cite{more1978levenberg}.

Given a nonlinear least squares problem:
\begin{equation}
\min_{x \in \mathbb{R}^n} \quad \frac{1}{2} \|f(\mathbf{x})\|_2^2,
\end{equation}
with Jacobian matrix \( J(\mathbf{x}) \in \mathbb{R}^{m \times n} \), the trust-region step \( p \in \mathbb{R}^n \) is constrained by a scaled norm condition:
\begin{equation}
\| D p \|_2 \leq \Delta,
\end{equation}
where \( D \in \mathbb{R}^{n \times n} \) is a diagonal scaling matrix and \( \Delta>0 \) is the \emph{trust-region radius}, i.e., the maximum allowed step length in the scaled variable space.

In the Jacobian-based strategy, the diagonal entries of \( D \) are defined from the column 2-norms of the Jacobian,
\begin{equation}
d_j = \| J_{:,j} \|_2,
\end{equation}
with the convention that \( d_j = 1 \) if the column norm is zero. These values are first set from the initial Jacobian and then updated monotonically at each iteration according to
\begin{equation}
d_j \;\leftarrow\; \max\{d_j, \|J_{:,j}\|_2\},
\end{equation}
so that the scaling can only increase as the algorithm proceeds. This guards against vanishing scales and ensures that once a variable is identified as sensitive, its step size remains controlled in later iterations.

The resulting scaled norm used in the trust-region constraint is
\begin{equation}
\| D p \|_2 = \sqrt{ \sum_{j=1}^n d_j^2 p_j^2 }.
\end{equation}

This scaling helps regularize the trust-region subproblem by approximately balancing the contributions of each variable based on their local sensitivity to the residuals. In practice, it can significantly improve numerical conditioning and convergence for problems where the variables differ in magnitude or physical units.

However, in the context of stellarator boundary optimization, this form of Jacobian-based scaling was insufficient to address the extreme dynamic range inherent in the Fourier representation of the boundary shape. In practice, the \texttt{scale='jac'} method did outperform the unscaled case (\texttt{scale=1}), but the improvement was limited and it consistently underperformed relative to the proposed ESS. The cross-sections of the stalled optimization when only using Jacobian scaling is shown in Figure~\ref{fig:failed_cross-sections}. By contrast, the solutions with ESS, shown in Figure~\ref{fig:ESS_cross-sections}, achieve superior results by explicitly aligning the scaling with the expected spectral decay of physically realistic boundary shapes.

\bibliographystyle{jpp}

\bibliography{ESS}

\end{document}